\documentclass[a4paper,12pt]{article}
\pdfoutput=1 

\usepackage{jcappub} 

\usepackage{amsmath}
\usepackage{amsfonts,amssymb}
\usepackage{hyperref}
\usepackage{graphicx}
\usepackage[utf8]{inputenc}
\usepackage{xcolor}
\usepackage{bm}
\usepackage{multirow}
\usepackage{diagbox}
\usepackage{hhline}

\usepackage[normalem]{ulem}

\hypersetup{
    colorlinks, 
    linktoc=all, 
    linkcolor=blue,  
  citecolor=hyptxt,
  urlcolor=blue}
  \hypersetup{
  citebordercolor=,
  filebordercolor=green,
  linkbordercolor=blue
}
\definecolor{hyptxt}{rgb}{0.7, 0.4, 0.9}

\newlength{\myleftlen}

\newcommand{\Vext}{\ensuremath{\textit{\textbf{V}}_{\text{ext}}}}
\title{\boldmath Cross-correlating radial peculiar velocities and CMB lensing convergence}

\author[a]{Leonardo Giani,}
\author[a]{Cullan Howlett,}
\author[a]{Rossana Ruggeri,}
\author[b,c,d]{Federico Bianchini,}
\author[a]{Khaled Said,}
\author[a]{Tamara M.~Davis}
\affiliation[a]{The University of Queensland, School of Mathematics and Physics,\\ QLD 4072, Australia}

\affiliation[b]{Kavli Institute for Particle Astrophysics and Cosmology,\\
Stanford University, 452 Lomita Mall, Stanford, CA, 94305, USA}
\affiliation[c]{SLAC National Accelerator Laboratory,2575 Sand Hill Road, Menlo Park, CA, 94025, USA}
\affiliation[d]{Department of Physics, Stanford University, 382 Via Pueblo Mall, Stanford, CA, 94305, USA}

\emailAdd{uqlgiani@uq.edu.au}
\emailAdd{c.howlett@uq.edu.au}
\emailAdd{r.ruggeri@uq.edu.au}
\emailAdd{k.saidahmedsoliman@uq.edu.au}
\emailAdd{fbianc@stanford.edu}
\emailAdd{tamarad@physics.uq.edu.au}

\abstract{
We study, for the first time, the cross correlation between the angular distribution of radial peculiar velocities (PV) and the lensing convergence of cosmic microwave background (CMB) photons.
We derive theoretical expectations for the signal and its covariance and assess its detectability with existing and forthcoming surveys. We find that such cross-correlations are expected to improve constraints on different gravitational models by partially breaking degeneracies with the matter density. We identify in the  distance-scaling  dispersion of the peculiar velocities the most relevant source of noise in the cross correlation. For this reason, we also study how the above picture changes assuming a redshift-independent scatter for the PV, obtained for example using a reconstruction technique. Our results show that the cross correlation might be detected in the near future combining PV measurements from DESI and the convergence map from CMB-S4. Using realistic direct PV measurements we predict a cumulative signal-to-noise ratio of approximately $3.8 \sigma$ using data on angular scales $3 \leq \ell \leq 200$.  For an idealized reconstructed peculiar velocity map extending up to redshift $z=0.15$ and a smoothing scale of $4$ Mpc $h^{-1}$ we predict a cumulative signal-to-noise ratio of approximately $ 27 \sigma$ from angular scales $3 \leq \ell \leq200 $. We conclude that currently reconstructed peculiar velocities have more constraining power than directly observed ones, even though they are more cosmological-model dependent. 
}

\begin{document}
\maketitle
\flushbottom
\section{Introduction}

Nowadays, at least on cosmological scales, the most commonly accepted description for the evolution of the Universe is the $\Lambda$CDM model~\cite{Planck:2018vyg,DES:2021wwk,PhysRevD.103.083533,Brout:2022vxf}, which consistently accounts for a smooth transition between three distinct stages of evolution. More precisely, it describes an expanding Universe which is initially filled mostly by radiation, i.e. photons and neutrinos. The latter are diluted to the point that their density becomes comparable to or smaller than the Cold Dark Matter (CDM) one. At this stage the photons are initially confined by the presence of baryons in a dense plasma, but eventually become free once that the Universe expanded enough to make the Thomson scattering inefficient. The last photons that scattered off the electrons constitute the \textit{Cosmic Microwave Background} (CMB). The initially small density and temperature fluctuations of matter and radiation are correlated, and trigger the gravitational collapse of matter. Hotter regions become denser and and colder ones emptier. Eventually, this led to the formation of a complex web of CDM structures where baryons are trapped, and undergo a similar process resulting in the formation of stars and galaxies. When the cosmic expansion dilutes the CDM density enough, the most abundant component of the Universe becomes the Cosmological Constant $\Lambda$. Maybe  coincidentally~\cite{PhysRevLett.82.896,Velten:2014nra}, this happens about half the age of the universe ago. As the name itself suggests, $\Lambda$ does not dilute and we enter the final stage of the evolution of the Universe.

From an observational point of view, it is relatively simple to observe the leading character of the first act in the $\Lambda$CDM expansion history by looking at the CMB. These photons travelled from their last Thomson scattering until today, and their distribution has been probed by space-based experiments like WMAP and Planck \cite{2013ApJS..208...19H,Planck:2018vyg}, and ground-based like the South Pole Telescope (SPT)~\cite{Carlstrom_2011} and the Atacama Cosmology 
Telescope (ACT) ~\cite{ACT:2020frw}. On the other hand, it is impossible to probe with direct observations the distribution of CDM and $\Lambda$ since they only interact through gravity. However, baryonic structures are formed within larger CDM haloes \cite{Angulo:2013qp}, and hence trace the underlying CDM distribution. The CMB photons are also sensitive to the matter distribution because of the gravitational lensing induced on them by the Large Scale Structure (LSS), hence the CMB itself is also a CDM tracer. This, in turn, suggests that the signals of the CMB and of any tracer of the CDM distribution should be correlated. This hypothesis was tested (and confirmed) using a variety of CDM tracers such as the spatial distributions of galaxies, Type Ia supernovae, and quasars, see for example Refs.~\cite{DES:2022ign,DES:2015eqk,DES:2018miq,PhysRevD.86.083006,Singh:2016xey,HerschelATLAS:2014txv}.

It is important to stress that these tracers are non trivially related to the underlying total matter distribution $\delta =(\rho-\bar{\rho})/\bar{\rho}$, where $\rho$ is the total matter density and $\Bar{\rho}$ its mean. Indeed, one needs to introduce auxiliary parameters not predicted by the $\Lambda$CDM model itself, like the galaxy bias $b$ relating the CDM density to the number density of galaxies. These auxiliary parameters and the cosmological ones may be (and often are) degenerate in the actual measurement, leading to potential model dependent biases in the cosmological inference. Luckily, the self-correlation  and the  cross correlations of different tracers may break these degeneracies, making this subject an active field of research ~\cite{Fabbian:2019tik,Pearson:2013iha}. Furthermore, over the last decade, high precision cosmology measurements have challenged our understanding of the growth and distribution of LSS due to an established $\sim 3\sigma$ tension on the observed clustering rate at the pivot scale of 8 Mpc between early and late Universe cosmological probes, usually referred to as $S_8$ tension \cite{DiValentino:2020vvd,Benisty:2020kdt,Kazantzidis:2019nuh}. This tension, interestingly, is closely related to the one on the value of the Hubble factor today $H_0$ \cite{Bernal:2016gxb,Verde:2019ivm,DiValentino:2020zio}, as it is difficult to alleviate either of the two without worsening the other \cite{Abdalla:2022yfr,Bhattacharyya:2018fwb,Heisenberg:2022gqk}. These challenges to the standard model paradigm advocate for new independent probes of the LSS distribution. 

With the above motivations in mind, in this work we investigate for the first time the cross-correlation between the CMB lensing signal and the distribution of radial peculiar velocities (PV) of galaxies along the line of sight. Indeed, due to the presence of LSS,  galaxies are not at rest with respect to the cosmological rest frame and move from underdense regions to overdense ones, following the strength of the local gravitational potential. We define the radial PV as the projection along the line of sight of the  velocity induced by these inhomogeneities. In recent years, several works have shown \cite{Koda:2013eya,Huterer:2016uyq,Macaulay:2016uwy,Andersen:2016ywu,Howlett:2017asw,Amendola:2019lvy,Quartin:2021dmr,Whitford:2021xqk,Lai:2022sgp} the positive impact of PV measurements in combination with redshift survey data in constraining cosmological
parameters. Ongoing and upcoming surveys like DESI \cite{DESI:2016fyo} and 4HS \cite{https://doi.org/10.18727/0722-6691/5117} will provide an order of magnitude more PV measurements than existing catalogs like SDSS \cite{Howlett:2022len}, with increased precision and redshift depth. For these reasons, cosmological inference with PV is an active field of research and a promising avenue to test our understanding of gravity.

At linear order in perturbation theory, the gradient of the PV is proportional to the matter density contrast $\delta$, making them a powerful tracer of the CDM distribution with no dependence on the galaxy bias $b(z)$. 
State of art PV surveys usually target objects located within redshift $z \leq 0.1$. On the other hand, most of the gravitational lensing is induced by the LSS at higher redshift, $z\sim 1$. Because of this huge difference, one would naively expect the signal from the cross correlation of the two to be essentially vanishing. Surprisingly, as we are going to show, this is not the case. Even if intrinsically small, our results indicate that current and forthcoming experiments may be able to detect the aforementioned cross-correlation. Our findings show that the main obstacle towards the detection of the signal is the scale-dependent dispersion associated with direct PV measurements. We also find that reconstructed PV can alleviate this problem, but require model dependent assumptions on cosmological parameters such as the linear galaxy bias $b(z)$.

The structure of the paper is the following: in Sec.~\ref{angcorr} we briefly review the formalism used to define angular cross-correlations between cosmological observables. In Sec.~\ref{PVCMB-theory} we compute the theoretical expectation for the cross-correlation between the CMB convergence $\kappa$, a measurement of the CMB lensing, and radial PV as function of the angular scale $\ell$. We also discuss the physical properties of the signal, and derive the covariance matrix. In Sec.~\ref{Dataset} we introduce the datasets used in this work, and in Sec.~\ref{Results} we present our results for the expected cumulative signal-to-noise ratio (S/N)  for existing and forthcoming surveys. Finally, in Sec.~\ref{Discussion} we discuss our findings and address future shortcomings.


\section{Theory of angular correlations for Cosmological observables}\label{angcorr}
In this section, following closely Ref.~\cite{Schoneberg:2018fis}, we briefly review the theory of angular statistics for cosmological observables. To begin with, we note that any observable $\mathcal{O}^\alpha(\hat{\textbf{n}},z)$ can be expanded in spherical harmonics with respect to its angular position on the sky $\hat{\textbf{n}}$:
\begin{equation}\label{angularexpreal}
    \mathcal{O}^\alpha(\hat{\textbf{n}},z)= \sum_{\ell m}Y_{\ell m}(\hat{\textbf{n}})a_{\ell m}^\alpha(z)\;, \qquad a_{\ell m}^{\alpha}(z)=\int d\Omega_{\hat{\textbf{n}}} \; Y^{*}_{\ell m}(\hat{\textbf{n}})\mathcal{O}^\alpha(\hat{\textbf{n}},z) \;.
\end{equation}
The correlation function between the spherical harmonic coefficients $a_{lm}^\alpha$'s defines the angular cross spectrum,
\begin{equation}
    C_\ell^{\alpha\beta}(z_1,z_2)\equiv \left<a_{\ell m}^{\alpha,*}(z_1),a_{\ell m}^\beta(z_2)\right>\;.
\label{eq:clm}
\end{equation}
On the other hand, our theoretical predictions are usually derived from some complicated system  of differential equations which are often more conveniently handled in Fourier space rather than in real space. The Fourier transform of $\mathcal{O}$ reads,
\begin{equation}
    \mathcal{O}^\alpha(\hat{\textbf{n}},z) = \int \frac{d^3 k}{\left(2\pi\right)^3} e^{i \textbf{k}\cdot\textbf{x}}\mathcal{O}^{\alpha}(\textbf{k},z)\;.
\end{equation}
The plane waves may be rewritten using the spherical harmonics addition theorem,
\begin{equation}
    e^{i \textbf{k}\cdot\textbf{x}} = 4\pi\sum_{\ell=0}^{
\infty}\sum_{m=-\ell}^\ell i^l j_\ell\left(kx\right)Y_{\ell m}(\hat{\textbf{k}})Y^{*}_{\ell m}\left(\hat{\textbf{n}}\right) \;,
\end{equation}
where the $j_\ell$ are the spherical Bessel functions of the first kind.
After substitution in Eq.~\eqref{angularexpreal}, exploiting the orthogonality of the spherical harmonics, we finally obtain
\begin{equation}
    a_{\ell m}^\alpha(z) = 4\pi i^\ell\int \frac{d^3k}{\left(2\pi\right)^3} j_\ell\left(k\chi(z)\right)\mathcal{O}^{\alpha}(\textbf{k},z)Y_{\ell m}^*(\hat{\textbf{k}})\;.
\label{eq:alm}
\end{equation}
For cosmological purposes, all the observables of interest can be split into a time and scale (and observable) dependent transfer function $T^\alpha(k,z)$ and a  primordial  perturbation $\mathcal{R}(\textbf{k})$ as $\mathcal{O}^{\alpha}(\textbf{k},z) = T^{\alpha}(k,z)\mathcal{R}(\textbf{k})$. We assume that the primordial perturbation is  Gaussian~\cite{Bardeen:1983qw,Lyth:1984gv,1990MNRAS.244..188B} and satisfies 
\begin{equation}
    \left<\mathcal{R}(\textbf{k}),\mathcal{R}(\textbf{k}')\right>=\frac{2\pi^2}{k^3}P_{\mathcal{R}}(k)\delta\left(\textbf{k}-\textbf{k}'\right)\;,
\label{eq:pkr}
\end{equation}
where $P_{\mathcal{R}}(k)$ is the power spectrum of the primordial perturbation. The above expression implicitly assumes that the primordial perturbation is adiabatic, i.e. induces inhomogeneities in the spatial curvature. This assumption is reasonable in light of the Planck constraints on entropic perturbations \cite{Planck:2018jri}, even though certain forms of isocurvature perturbations, for example compensated ones, are not well constrained by Planck \cite{Barreira:2023uvp}.
Note that by construction we are assuming that different modes $\textbf{k},\textbf{k}'$ are independent of each other. 

Substituting expressions \ref{eq:alm} and \ref{eq:pkr} into \ref{eq:clm} we arrive at
\begin{equation}
    C_\ell^{\alpha\beta}(z_1,z_2) = 4\pi\int_0^\infty \frac{dk}{k} P_{\mathcal{R}}(k)T^{\alpha}(k,z_{1}) T^{\beta}(k,z_{2}) j_\ell(k\chi(z_{1}))j_\ell(k\chi(z_{2}))\;.
\end{equation}
In actual measurements,  rather than $\mathcal{O}$, one usually observes its average over a redshift bin weighted by some kernel $W(z)$. Accordingly, this can be absorbed in to the definition of the angular power spectrum
\begin{equation}
    C_\ell^{\alpha\beta} = \int W(z_{1})W(z_{2})C_\ell^{\alpha\beta}(z_1,z_2) dz_{1} dz_{2} = 4\pi\int_0^\infty \frac{dk}{k} P_{\mathcal{R}}(k)\Delta^\alpha_\ell(k)\Delta^\beta_\ell(k)\;,
\end{equation}
where the kernel  $\Delta^\alpha_\ell$ is defined as 
\begin{equation}\label{kerneldef}
    \Delta^\alpha_\ell (k) \equiv \int_0^{\infty}dz\, T^{\alpha}(k,z) j_\ell(k\chi(z)) W^{\alpha}(z)\;. 
\end{equation}

\section{Cross correlation between CMB lensing and Peculiar Velocities}
\label{PVCMB-theory}
The general derivation in the previous section worked primarily with redshifts, as this is the observable property of galaxy surveys. However, for the purposes of both CMB lensing and peculiar velocity theory, it is useful to consider distances. As such, for the remainder of this work we provide expressions in terms of the comoving distance
\begin{equation}
    \chi = \frac{c}{H_0}\int_{0}^z  \frac{dz}{E(z)}\;, \qquad E(z) = \frac{H(z)}{H_0}\;, \quad  H(z) = H_0\sqrt{\Omega_m(z) + \Omega_r(z) + \Omega_\Lambda}\;,
\end{equation}
where $\Omega_m$ and $\Omega_{\Lambda}$ are the matter and cosmological constant energy densities and where $H_0 = H(z=0)$ is the value of the Hubble factor today.
The necessary factors for converting between integration over $z$ and integration over $\chi$ have generally been absorbed into the definition of the window function $W(\chi)$ below.

\subsection{Theoretical prediction}
The Einstein field equations for scalar perturbations of a flat FLRW background in the Newtonian Gauge give, at linear order, the following relation between the peculiar velocity field $v(\textbf{r})$ and the dark matter density contrast $\delta(\textbf{r})$ at position $\textbf{r}$:
\begin{equation}\label{linearvel}
    \nabla \cdot v(\textbf{r}) = aHf\delta(\textbf{r})\;,
\end{equation}
where $a$ is the scale factor, $H$ the Hubble factor and $f = d\ln D/d\ln a$ the growth rate,  which is itself a function of the growth function $D$,  the growing mode of the linear density contrast.
These are in general functions of the redshift, but in the following we will omit their explicit dependence.
In Fourier space, the relation between velocity and density becomes
\begin{equation}
    v(\textbf{k}) = -iaH f \frac{\delta(\textbf{k})}{\textbf{k}}\;.
\end{equation}
In current practical experiments, however, one can usually measure  only the radial projection of the velocity field, 
\begin{equation}
    u(\hat{\textbf{r}},\textbf{k}) \equiv v(\textbf{k})\cdot \hat{\textbf{r}} = -iaHf\frac{\delta\left(\textbf{k}\right)}{k^2}\textbf{k}\cdot\hat{\textbf{r}}\;,
\end{equation}
where $k \equiv |\textbf{k}|$ and $\hat{\textbf{r}} \equiv \textbf{r}/|\textbf{r}|$. The Fourier transform of this reads
\begin{equation}
u(\textbf{r})= -i\int \frac{d^3k}{(2\pi)^3}aHf\;\frac{\delta\left(\textbf{k}\right)}{k^2}\textbf{k}\cdot\hat{\textbf{r}}\;e^{i\textbf{k}\cdot\textbf{r}}\;.
\end{equation}
The above expression may be simplified, noticing that
\begin{equation}
    \hat{\textbf{k}}\cdot\hat{\textbf{r}}\;e^{i\textbf{k}\cdot\textbf{r}} = -i\frac{d}{d(kr)}e^{i\textbf{k}\cdot\textbf{r}}\;,
\end{equation}
from which, using again the spherical harmonics addition theorem, the spherical harmonic coefficients for the radial velocity field may be written as
\begin{equation}
    a^{u}_{\ell m}(\chi) = (4\pi i^\ell)aHf\int\frac{d^3k}{(2\pi)^3}\frac{\delta(\textbf{k})}{k}j'_\ell\left(k\chi\right)Y^*_{\ell m}(
    \hat{\textbf{k}})\;.
\end{equation}
Projecting the above along the line of sight we may define the radial velocity kernel $\Delta^{u}_\ell(k)$ as
\begin{equation}
\label{pvker}
        \Delta^{u}_\ell(k)  \equiv \frac{1}{k}\int_0^{\infty}d\chi\; W^{u}(\chi) j'_\ell\left(k\chi\right) D(\chi)\;,
\end{equation}
where we have introduced the window function $W^{u}(\chi)$,
\begin{equation}
    W^{u}(\chi) = Hfa\frac{dn}{d\chi}\;,
\end{equation} 
and  the distribution of sources $dn/d\chi$. For comparison, the kernel for the galaxy distribution, neglecting the magnification bias, can be written
\begin{equation}\label{gker}
    \Delta^{g}_\ell(k) \equiv \int_0^{\infty}d\chi\; W^{g}(\chi) j_\ell\left(k\chi\right) D(\chi) = \int_0^{\infty}d\chi\;b(\chi) \frac{dn}{d\chi} j_\ell(k\chi)D(\chi) \;,
\end{equation}
where $b(\chi)$ is the linear galaxy bias. 
The CMB lensing signal, following Ref.~\cite{DES:2015eqk}, can be expressed in terms of the convergence field $\kappa(\hat{\textbf{n}})$ in the direction $\hat{\textbf{n}}$:
\begin{equation}
    \kappa(\hat{\textbf{n}})= \frac{3}{2c^2}\Omega_m H_0^2 \int_0^{\chi_{\rm CMB}} d\chi \frac{\chi}{a(\chi)}\frac{\chi_{\rm CMB}-\chi}{\chi_{\rm CMB}}\delta\left(\chi \hat{\textbf{n}}, \eta_0-\chi\right)\;,
\end{equation}
where $c$ is the speed of light and $\chi_{\rm CMB}$ the comoving distance to the surface of last scattering.
From the above we can define the convergence kernel $\Delta^\kappa_\ell(k)$ and the convergence projection kernel $W^{\kappa}(\chi)$ as:
\begin{equation}\label{kappaker}
    \Delta^\kappa_\ell (k) \equiv  \int_0^{\chi_{\rm CMB}}d\chi\; W^{\kappa}(\chi) j_\ell\left(k\chi\right) D(\chi)\;,
\end{equation}
\begin{equation}
   W^{\kappa}(\chi) \equiv \frac{3}{2c}\Omega_m H_0^2\; \chi\frac{1+z\left(\chi\right)}{H\left(\chi\right)}\frac{\chi_{\rm CMB}-\chi}{\chi_{\rm CMB}}\;.
\end{equation}

Combining expressions \eqref{pvker} and \eqref{kappaker}, we finally obtain an expression for their angular cross correlation
\begin{equation}\label{CCPVCMB}
    C_\ell^{u\kappa} \equiv \left<u^*,\kappa\right>     = \frac{2}{\pi}\int dk P(k)k^2 \Delta^u_\ell(k)\Delta^\kappa_\ell(k)\;, 
    \end{equation}
    which in terms of the window functions $W^i$ becomes, 
    \begin{equation}\label{CCPVCMBWF}
C_\ell^{u\kappa} \equiv \frac{2}{\pi}\int dk\;d\chi_1\;d\chi_2\;k\;W^{u}(\chi_1)W^{\kappa}(\chi_2) j'_\ell\left(k\chi_1\right) j_\ell\left(k\chi_2\right)P_m\left(k,\chi_1,\chi_2\right)\;.  
\end{equation}
Note that since we are restricting our analysis to linear scales where Eq.~\eqref{linearvel} is a valid approximation, we have defined the matter power spectrum $P_m\left(k,\chi_1,\chi_2\right) \equiv P(k)D(\chi_1)D(\chi_2)$, with $P(k) = (k^{3}/2\pi^2)P_{\mathcal{R}}(k)T(k)^2$.\footnote{In other words we are assuming that the transfer function appearing in Eq.\eqref{kerneldef} is well approximated by the product  of the growth function $D(z)$ with a time-independent transfer function $T(k)$.}

Looking at the cosmological parameters entering in the kernels for the galaxy density, velocity and CMB convergence, we can see that the benefit of measuring the cross-correlation $C^{uk}_{\ell}$ is that it provides additional constraining power on the growth rate of structure compared to measuring the auto-correlation of the PVs, $C^{uu}_{\ell}$, alone. It is also independent of galaxy bias.\footnote{In this work we have neglected redshift-space distortions (RSD) and so the galaxy-density kernel contains only $b(z)$ at linear order. If we were to include linear-scale RSD \cite{Kaiser1987} then we would see that $C^{gg}_{\ell}$ also contains information on the growth rate of structure, however at the same order and hence degenerate with the galaxy bias.} Furthermore, as the CMB-lensing kernel contains information on the matter density, this may allow one to partially break the degeneracies between different models of gravity, with different growth rates of structure, and the matter density. This is simplest to see if we consider the common `$\gamma$' parameterisation of the growth rate, $f(z) = \Omega_{m}^{\gamma}(z)$, where $\gamma = const \approx 0.55$ for General Relativity, but differs in other gravitational theories \cite{Linder2005,Linder2007}. Including measurements of $C^{uk}_{\ell}$ alongside the auto-correlations of the velocity or CMB-convergence clearly provides a route to further break the degeneracy between $\Omega_{m}$ and $\gamma(z)$ in the above expression.

\subsection{Peeking ``beyond'' the survey}
\begin{figure}[h]
\includegraphics[scale=0.5,trim=0mm 00mm 0mm 0mm]{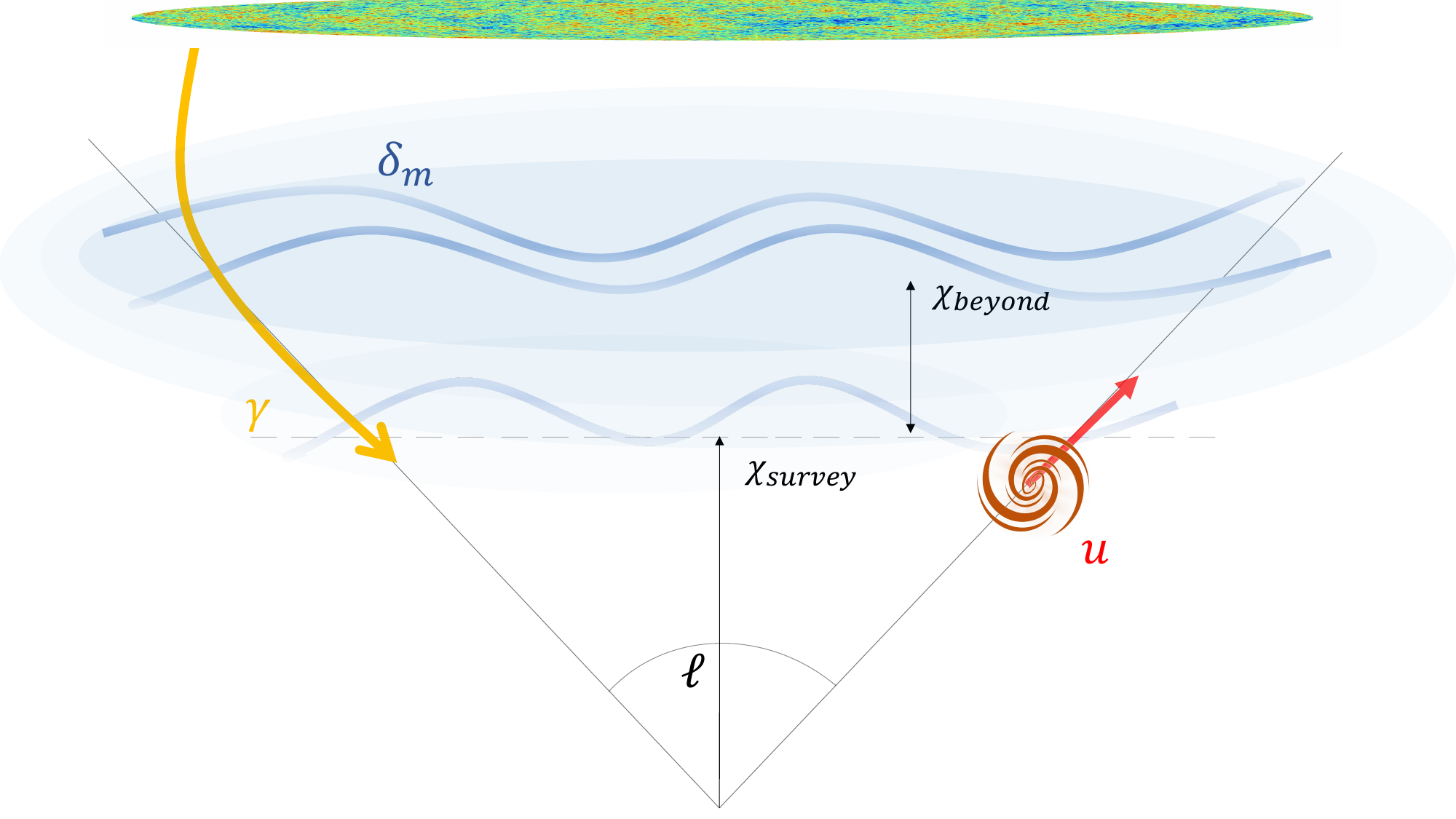}\;
	\caption{Schematic representation of a lensed CMB photon $\gamma$ (in yellow) and of the radial peculiar velocity $u$ (in red) of a galaxy at the edge of the PV survey, i.e. at a distance $\chi_{\rm survey}$, separated by an angular scale $\ell$.  According to Eq.~\eqref{CCPVCMBLimber}, their correlation includes a contribution from the matter density distribution (in blue) at a distance $\chi_+ = \chi_{\rm survey}+\chi_{\rm beyond}$, where $\chi_{\rm beyond}/\chi_{\rm survey}\approx \left(\ell -\frac{1}{2}\right)^{-1}$. }
	\label{pvpeeking}
\end{figure}

An interesting feature of the radial PV field is that it is  effectively sensitive to the CDM distribution outside the scope 
of the PV survey itself. The same is true for its cross correlation with the convergence field $\kappa$, as reflected by the derivatives of the spherical Bessel function appearing in Eq.~\eqref{CCPVCMB}.
This and other features of the $C^{u\kappa}_{\ell}$  are more easily understood, at qualitative level, estimating Eq.\eqref{CCPVCMBWF} using the Limber approximation to compute the integrals over $\chi_1$ and $\chi_2$ (see Appendix \ref{appendixA} for the  details of the derivation):
\begin{equation}\label{CCPVCMBLimber}
\begin{split}
    C^{u\kappa}_{\ell} \approx \int \frac{dk}{k} &\sqrt{\frac{1}{\ell+\frac{1}{2}}}   \left[\frac{W^{u}\left(\frac{\ell-\frac{1}{2}}{k}\right)W^\kappa\left(\frac{\ell+\frac{1}{2}}{k}\right)}{\sqrt{\ell -\frac{1}{2}}}P_m\left(k,\frac{\ell-\frac{1}{2}}{k},\frac{\ell+\frac{1}{2}}{k}\right)\right. \\
    &-\left. \frac{\left(\ell+1\right)W^u\left(\frac{\ell+\frac{1}{2}}{k}\right)W^{\kappa}\left(\frac{\ell+\frac{1}{2}}{k}\right)}{\sqrt{\ell+\frac{1}{2}}\left(\ell+\frac{1}{2}\right)}P_m\left(k,\frac{\ell+\frac{1}{2}}{k},\frac{\ell+\frac{1}{2}}{k}\right)\right]\;.
\end{split}
\end{equation}
As we are going to show in Sec.~\ref{Results}, most of the signal in $C^{u\kappa}_{\ell}$ comes from large angular scales, i.e. low-$\ell$ modes, for which the Limber approximation is inaccurate. With the exquisite precision expected from forthcoming cosmological surveys, alternative approximation schemes are currently developed in order to evaluate this type of integrals with greater accuracy, see for example Refs.~\cite{LSSTDarkEnergyScience:2022lno,Fang:2019xat}. However, all the result presented in this paper were obtained integrating numerically Eq.\eqref{CCPVCMBWF}, without the use of any approximation. On the other hand, a number of physical properties of the signal are straightforwardly drawn from Eq. \eqref{CCPVCMBLimber}. For example, one can appreciate that the first term within the square brackets combines the $\kappa$ and $u$ fields evaluated at  $\chi_{\pm}= (\ell \pm 1/2)/k$, which for a given $\ell$ are separated by a comoving distance proportional to $k^{-1}$. 
As an illustrative example, consider a galaxy in the north polar direction at a comoving distance $\chi_- \approx 300 h^{-1}\mathrm{Mpc}$, i.e. at the edge of current PV surveys. According to Eq.~\eqref{CCPVCMBLimber}, the radial peculiar velocity of this galaxy is correlated with the lensing induced on the trajectory of a CMB photon detected at the equatorial plane, i.e. for $\ell = 2$, due to the CDM distribution at a comoving distance of almost $\chi_+ \approx 500 h^{-1}\mathrm{Mpc}$.\footnote{We must stress that the above example is qualitative only and must be taken with grain of salt, as the error induced by the Limber approximation is proportional to $\ell^{-2}$, and is therefore of order $\sim 25\%$ for $\ell=2$.} Put another way, the same distribution of matter located beyond the edge of the survey is responsible both for the large-scale motions of galaxies within the survey \textit{and} the lensing of CMB photons, giving rise to a non-zero correlation. A schematic representation of the above effect is given in Fig. \ref{pvpeeking}.

\subsection{Overlap of the tracers}

\begin{figure}[h]
\centering
\includegraphics[scale=0.52, trim=20mm 0mm 0mm 5mm]{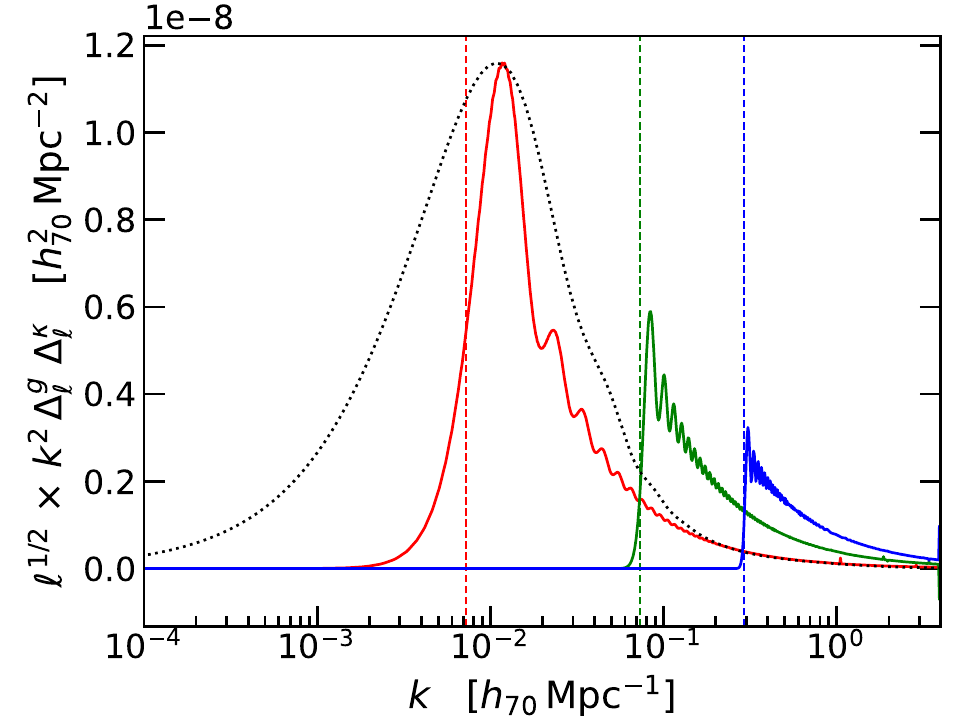}
\includegraphics[scale=0.52, trim=0mm 0mm 20mm 5mm]{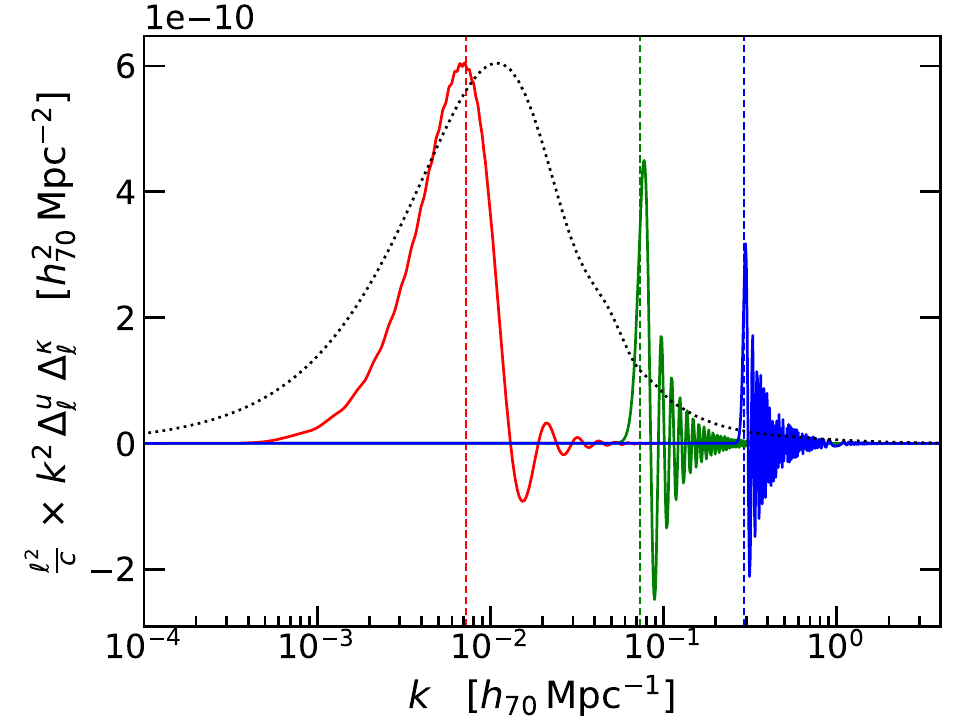}\\
\includegraphics[scale=0.52]{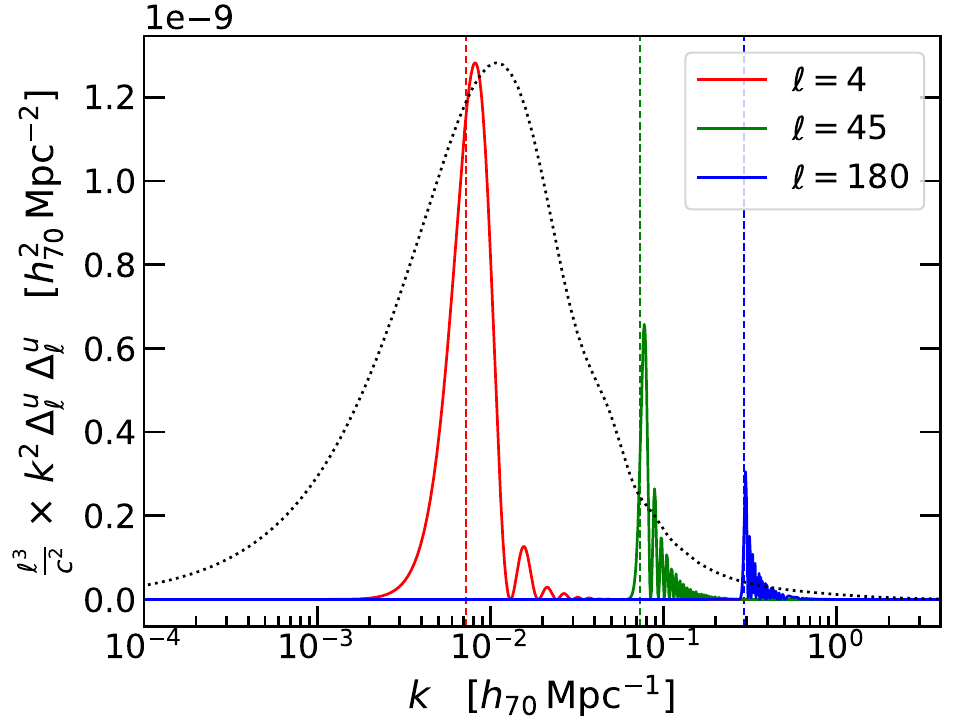}
	\caption{Weighting kernels as a function of scale $k$ for the angular power spectra of density-CMB convergence ($g$-$\kappa$), radial velocity-CMB convergence ($u$-$\kappa$) and radial velocity-radial velocity ($u$-$u$) are shown in the upper-left, upper-right and lower panels respectively, for a simple survey with a constant $1200$ galaxies per $0.001$ redshift bin (similar in scope to the DESI-PV survey). Red, green and blue lines correspond to the contribution of each $k$ mode to angular modes $\ell=4$, $45$ and $180$ respectively. The dashed vertical lines are the values of $\ell/\chi(z_{\rm max}=0.15)$ where $\chi(z_{\rm max}=0.15)$ is the maximum comoving distance probed by the simple survey. The dotted line is the linear matter power spectrum with arbitrary normalisation, which multiplies the weighting kernels when computing the $C_{\ell}$'s.}
	\label{deltas}
\end{figure}

We easily realize from Eq.~\eqref{CCPVCMB} that, for a given $\ell$, the amount of correlation between the $u$ and $\kappa$ fields is essentially given by the overlap of the two kernels $\Delta^{u}\Delta^{\kappa}$ weighted by the power spectrum $P(k)$. In  Fig.~\ref{deltas} we compare the product of these kernels with the galaxy-lensing and velocity-velocity ones, given as a function of the scale $k$ at fixed angular mode $\ell$ for an idealized uniform distribution of sources which extends up to a distance $\chi(z_{\rm max})$, with $\chi(z_{\rm max})$ being the maximum distance probed by the survey. Note that these curves peak around $k=\ell/\chi(z_{\rm max})$ and quickly decay for smaller values of $k$, as expected in virtue of the Limber approximation. Indeed, using the latter approximation the galaxy density kernel may be written:
\begin{equation}\label{approxdeltag}
    \Delta^{g}_\ell(k)\approx W^{g}\left(\frac{\ell+\frac{1}{2}}{k}\right) \;,  
\end{equation}
which for a uniform distribution of sources is roughly constant and non-vanishing only for $z\leq z_{\rm max}$, whereas the radial PV kernel may be written:
\begin{equation}\label{deltau}
    \Delta^{u}_\ell(k)\approx \frac{1}{\sqrt{\ell -\frac{1}{2}}}W^{u}\left(\frac{\ell-\frac{1}{2}}{k}\right) - \frac{\ell+1}{\sqrt{\ell+\frac{1}{2}}\left(\ell+\frac{1}{2}\right)}W^{u}\left(\frac{\ell+\frac{1}{2}}{k}\right)\;.   
\end{equation}
It is straightforward to realize that, since  the lensing convergence kernel $W^{\kappa}$ at small redshift grows roughly linearly (see the dashed line in Fig.~\ref{windows}), the product $\Delta^g_\ell\Delta^\kappa_\ell$ is maximum at $k = (\ell+\frac{1}{2})/\chi(z_{\rm max})$. A similar statement holds true for the product between the velocity  and lensing kernels. Indeed, the difference between the two terms in Eq.~\eqref{deltau} for large $\ell$ is always positive and close to zero unless $(\ell - \frac{1}{2})/\chi(z_{\rm max}) <k<(\ell + \frac{1}{2})/\chi(z_{\rm max})$, for which the second term is vanishing due to the lack of observed sources beyond $\chi(z_{\rm max})$. 

Notice that the decaying of the peaks at large scales, i.e. towards smaller $k$,  for the red curves in Fig.~\ref{deltas} is slower compared to the green and blue curves, and reflects the lack of precision of the Limber approximation used in Eqs.~\eqref{approxdeltag},\eqref{deltau} for small $\ell$. 
Another interesting feature  which characterizes the product $\Delta^\kappa_\ell \Delta^u_\ell$ is that, for a given $\ell$, it oscillates between positive and negative values. As one may deduce from Fig.~\ref{deltas}, its integral over $k$ is essentially dominated by the value of $\Delta^\kappa_\ell \Delta^u_\ell$  at $k = \ell /\chi(z_{\rm max})$. We also notice that, since the matter power spectrum is peaked around the scale of the horizon at matter-radiation equality, $k=k_{eq}$ and the amplitude of 
$\Delta^u_\ell$ for a uniform and constant distribution of sources\footnote{i.e for a distribution of sources such that, given the small redshift window considered here, $W\left(\frac{\ell + \frac{1}{2}}{k}\right)\approx W\left(\frac{\ell - \frac{1}{2}}{k}\right)$.} is a decreasing function of $\ell$, the correlation between the two fields is stronger at angular scales larger than $\ell \leq k_{eq}\chi(z_{\rm max})$.  On the other hand, at distances smaller than $\chi(z_{\rm max})$ the fluctuations of the radial velocity field average to almost zero. We conclude that most of the cosmological information from this correlation function comes from sources at the edge of the peculiar velocity survey, because of the lack of observed sources beyond it. For this reason, the expected signal is significantly smaller than, for example, the one from the galaxy density and CMB lensing cross correlation which contains contributions from all the sources within the galaxy survey.       

\subsection{Covariance}
To assess quantitatively the statistical significance of the cross correlation detection \eqref{CCPVCMB} with future realistic datasets we need to associate to it a noise, i.e. we need to build a covariance matrix. 
This is defined as the non-connected four-point correlation function
\begin{equation}
    \text{Cov}(C^{u\kappa}_{\ell},C^{u \kappa}_{\ell'})\equiv \Sigma^{u\kappa }_{\ell \ell'} = \left<a_{\ell m}^{\tilde{u},*},a_{\ell m}^{\tilde{\kappa}},a_{\ell' m'}^{\tilde{u}},a_{\ell' m'}^{\tilde{\kappa},*}\right>\;,
\end{equation}
where we have defined our measured radial velocity and convergence $\tilde{u},\tilde{\kappa}$ as
\begin{equation}
    \tilde{u} = u + \xi_{u}\;,
\end{equation}
\begin{equation}
    \tilde{\kappa} = \kappa + \xi_\kappa\;,
\end{equation}
with $\xi_{k,u}$ being the random (Gaussian) noises.
Expanding in spherical harmonics and making use of the Wick theorem (hence assuming that both $\kappa$ and $u$ are Gaussian fields) we can write the covariance as (see Eq.~15 of \cite{DES:2015eqk})
\begin{equation}\label{covar}
\Sigma^{u\kappa }_{\ell \ell'}= \frac{\delta_{\ell \ell'}}{f_{\rm sky}\left(2\ell + 1\right)}\left[\left(C^{\kappa \kappa}_{\ell} + N^{\kappa \kappa}_{\ell}\right)\left(C^{u u}_{\ell'}+N^{u u}_{\ell'}\right) + \left(C^{u\kappa}_{\ell}\right)^2\right]\;,                            
\end{equation}
where $f_{\rm sky}$ is the sky fraction where PV and CMB observations overlap, and $N^{\kappa\kappa}_{\ell}, N^{u u}_{\ell}$ are the noise spectra of the two fields.

We assume that the shot noise relative to the radial velocity field, being induced by the underlying galaxy distribution, is isotropic (does not depend on $l$) and can be estimated (see Appendix \ref{appendixb}) as
\begin{equation}
N^{uu}_{\ell} = \frac{1}{\bar{n}}\left(\int d\chi \frac{dn}{d\chi} (\alpha H_0 \chi) \right)^2 \;,
\label{eq:nuumain}
\end{equation}
where $\bar{n}$ is the angular number density of galaxies, and $\alpha$ is a scaling factor that depends of the type of tracer used for the peculiar velocity measurements. More details on these scalings are given in Section~\ref{Dataset}. For comparison, the usually assumed noise spectrum for the galaxy angular power spectrum is $N_{\ell}^{gg} = 1/\bar{n}$; the peculiar velocity field contains the same contribution from shot-noise as the galaxy density field, but has an additional term arising from the typical uncertainties in each PV measurement ($\alpha H_0 \chi$) weighted and integrated along the line-of-sight.

The noise power spectrum of the reconstructed CMB convergence field $\kappa$ depends non trivially on the specifics of the CMB experiment and on the particular reconstruction technique. 
The most commonly adopted method, the quadratic estimator, is based on the idea that lensing modifies the CMB statistics and induces a correlation between previously independent CMB harmonic modes \citep{Zaldarriaga:1998ar,Hu:2001kj}. 
By examining the correlation between modes of two CMB maps $X,Y \in T,E,B$, we can obtain a noisy estimate of $\phi^{XY}_{\ell m}=\frac{2}{\ell(\ell+1)}\kappa_{\ell m}^{XY}$ with associated noise power spectrum\footnote{The expression is strictly valid for the auto-spectrum of $\phi^{XY}$, i.e. $\langle \phi^{XY}\phi^{XY,*} \rangle$, for the general form see \cite{Okamoto:2003zw}.}
\begin{equation}
    N_L^{\phi\phi, XY} = L(L+1)(2L+1)\left[ \sum_{\ell\ell'} g_{\ell L \ell'}^{XY} f_{\ell L \ell'}^{XY}  \right]^{-1},
\end{equation}
where  $g_{\ell L \ell}^{XY} $ and $ f_{\ell L \ell}^{XY}$ are weight functions that depend on the power spectrum of the CMB fields $C_L^{XY}$ and their overall noise levels $N_{\ell}^{XY}$ (see \cite{Okamoto:2003zw} for the exact expressions).
The different temperature and polarization based estimators can be combined into a minimum-variance estimate, which is the nominal lensing reconstruction we assume for the forecasts presented in this work. 
The CMB lensing reconstruction noise curves for the surveys considered in this work, together with the signal power spectrum, are shown in Fig.~\ref{Cells}. 
Quadratic estimators will become sub-optimal at the instrumental noise levels soon-to-be reached by the most sensitive experiments, but a variety of methods based on maximum-likelihood and Bayesian techniques are being developed to restore optimality \citep[see, e.g.,][]{PhysRevD.96.063510,PhysRevD.102.123542,Bianchini:2022wte}.


\section{Peculiar velocity and CMB datasets}
\label{Dataset}
\begin{figure}[h]
\centering
\includegraphics[scale=0.8]{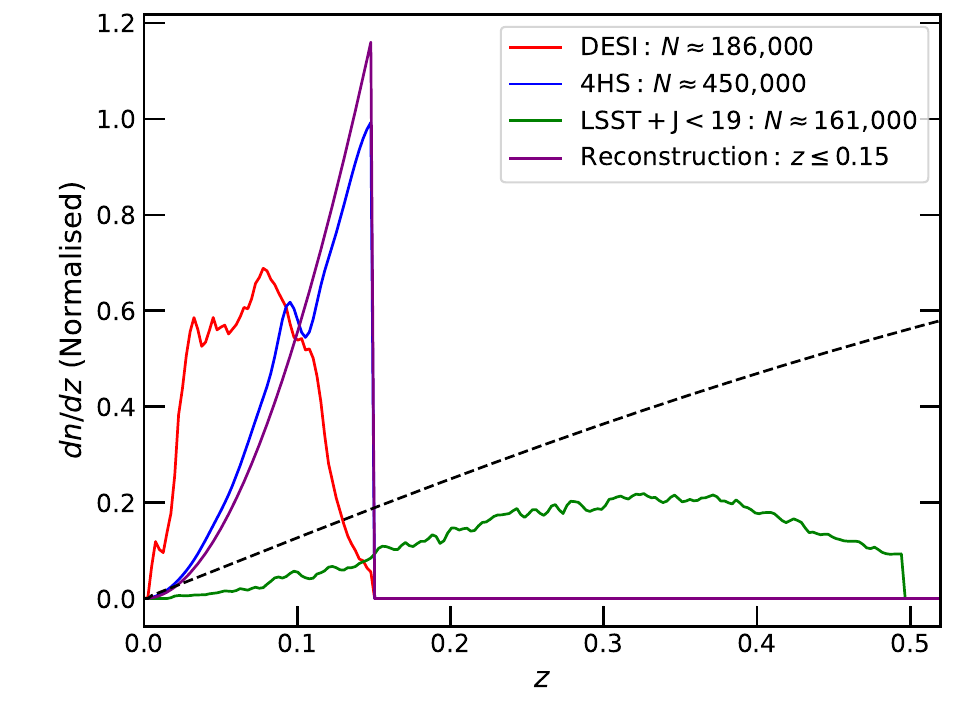}\;
	\caption{The distribution of sources nNormalised so that the integral of $dn/dz$ over all redshifts is one) for the four peculiar velocity datasets we consider in this work as a function of redshift. The dashed line is the CMB lensing kernel and the number of PV measurements for the three direct PV surveys we consider is given in the legend.}
	\label{windows}
 \end{figure}

\begin{figure}[h]
\centering
\includegraphics[scale=0.52, trim=20mm 0mm 0mm 5mm]{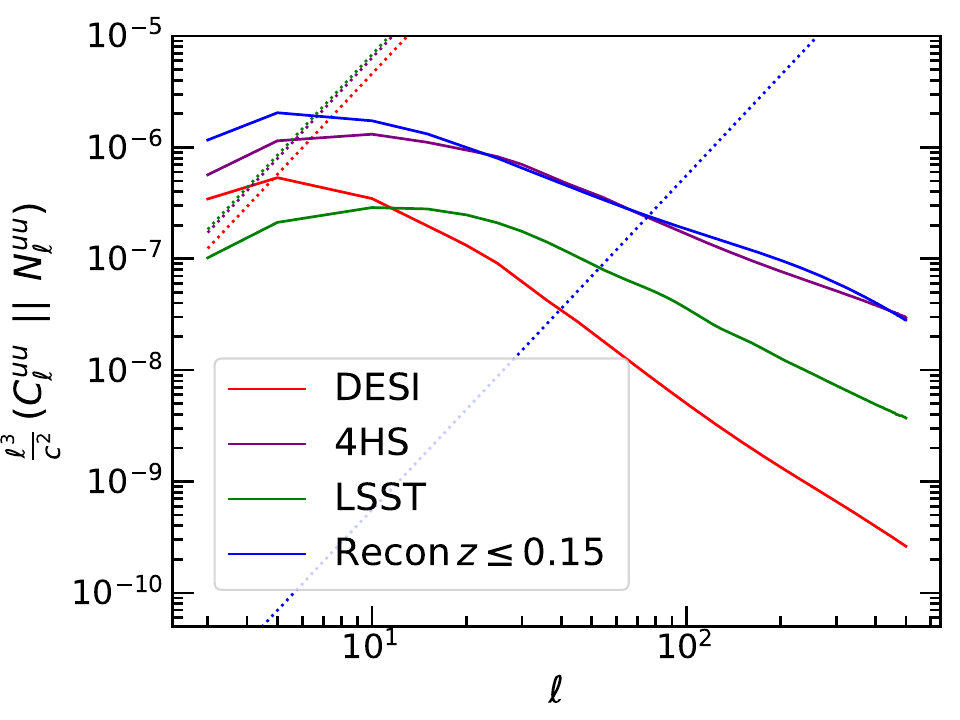}
\includegraphics[scale=0.52, trim=0mm 0mm 20mm 5mm]{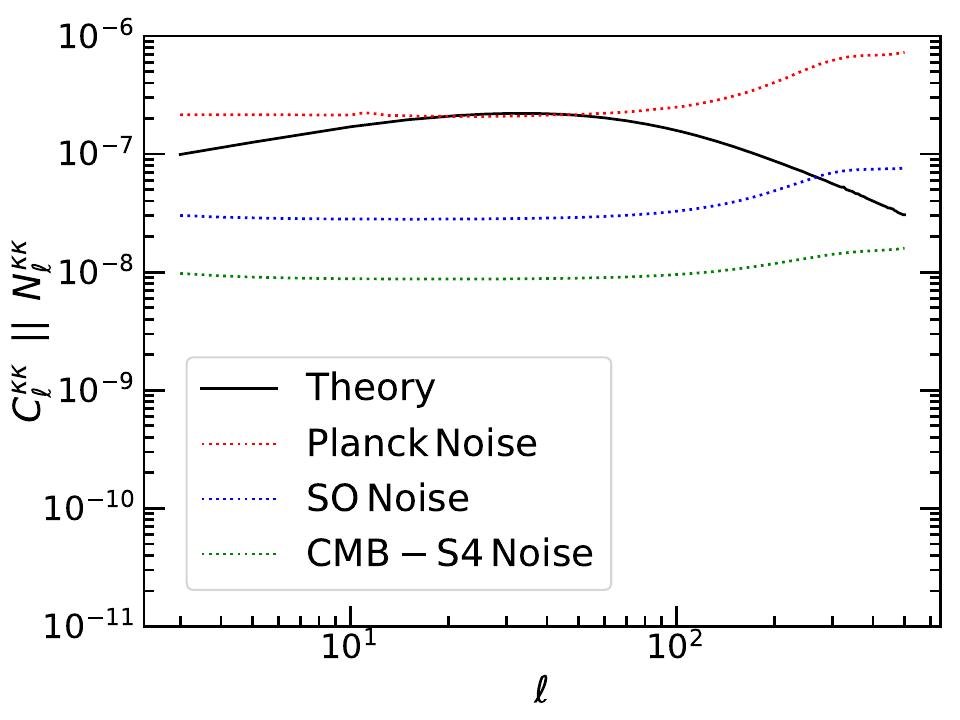}\\
\includegraphics[scale=0.52]{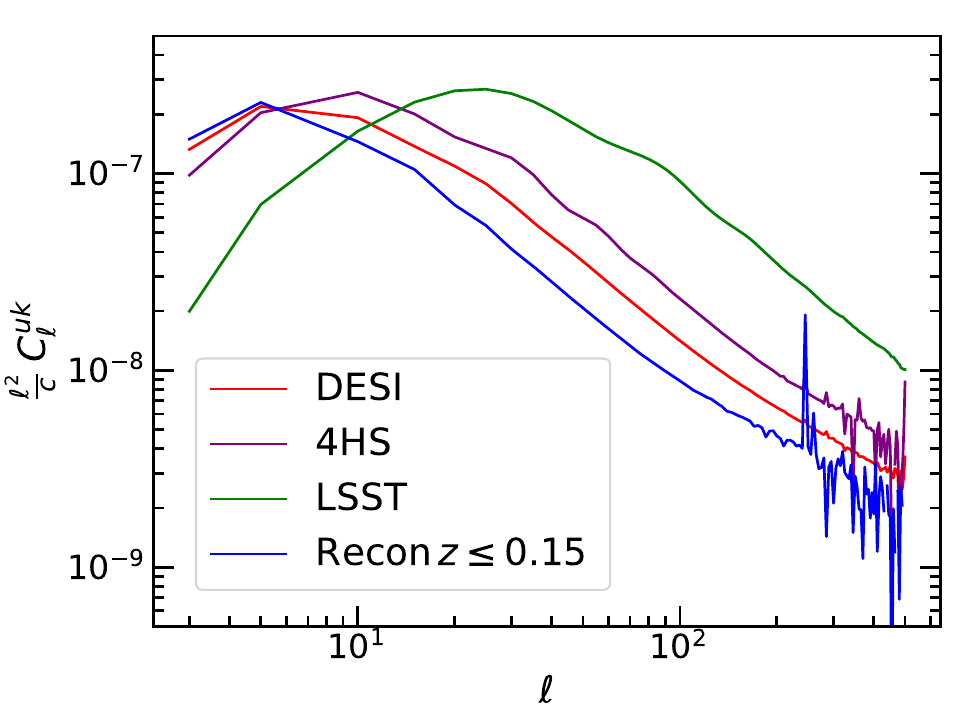}
	\caption{\textit{Upper left}: The angular power spectra and noises, solid and dotted lines respectively, of the radial velocity field for the various PV catalogs considered in this work. \textit{Upper right:} The angular power spectra and noises, solid and dotted lines respectively, of the lensing convergence field for the various CMB experiments considered in this work.\\
 \textit{Bottom}: The cross angular power spectrum of the velocity field and CMB lensing convergence for the different PV surveys considered in this work.}
	\label{Cells}
\end{figure}

So far we derived a theoretical prediction for the cross correlation signal and discussed its physical properties assuming idealized distributions of PV sources and lensed photons. To assess its potential as a cosmological probe, however, we should consider realistic datasets from existing and forthcoming experiments. In this section, we introduce the various PV and CMB catalogs we use to derive the results presented in Sec.~\ref{Results}. The number and the distribution of direct and reconstructed PV sources are summarized in Fig.~\ref{windows} and Table~\ref{PVsummary}. The signals and the noises for the radial velocity auto-correlation function and those of the various CMB lensing experiments  are reported in Fig.~\ref{Cells}.

\begin{table}[h]
\centering
\begin{tabular}{|c|cccc|}
\hhline{|=====|}

PV Survey & Area ($\mathrm{deg}^{2}$) & Depth ($z_{max})$ & $\Bar{n}$ (sr$^{-1}$) & $\alpha$ \\\hline
DESI  & 14000 & 0.1 & 5.9$\times 10^{4}$ & 0.2 \rule{0pt}{2.6ex} \\
4HS  & 17000 & 0.15 & 1.4$\times 10^{5}$ & 0.2 \\
LSST  & 18000 & 0.5 & 5.1$\times 10^{4}$ & 0.05 \\
Reconstruction   & Full Sky & 0.15 & 1.2$\times 10^6$ &  $(250\;\text{km s}^{-1})/(H_0\chi)$  \rule[-1.2ex]{0pt}{0pt} \\
\hhline{|=====|}
\end{tabular}
\caption{Summary statistics of the various PV catalogs used in this work. $\Bar{n}$ is the angular number density of sources per steradian, while $\alpha$ is the typical uncertainty on the peculiar velocities as a fraction of $H_{0}\chi$.}
\label{PVsummary}
\end{table}
 
\subsection{Direct PV catalogs}
\begin{enumerate}

\item[DESI] The Dark Energy Spectroscopic Instrument (DESI)\footnote{\href{https://www.desi.lbl.gov/}{https://www.desi.lbl.gov/}}~\cite{DESI:2016fyo} is a multi-fiber optical spectrograph  mounted on the Kitt Peak National Observatory Mayall 4m telescope. This instrument is being used to carry out a large survey of galaxy redshifts over 14000 deg$^2$ of sky, and over the full 5-year survey lifetime is also expected to collect velocities from $ \approx 180 \times 10^3$ elliptical and spiral galaxies up to 
$z\sim 0.1$ with an expected uncertainty which scales with the 20$\%$ of the distance to the target, i.e. such that $\alpha =0.2$ in Eq.~\eqref{eq:nuumain} (Saulder et. al., in prep.).  
\item[4HS] The 4MOST (4-metre Multi-Object
Spectroscopic Telescope) Hemispheric Survey (4HS)\footnote{\href{https://www.4most.eu/cms/science/extragalactic-community-surveys/}{https://www.4most.eu/cms/science/extragalactic-community-surveys/}} \cite{https://doi.org/10.18727/0722-6691/5117} is a forthcoming spectroscopic survey facility for the four-metre-class Visible
and Infrared Survey Telescope for Astronomy (VISTA) at the Paranal observatory, in Chile. Whilst having similar uncertainty to DESI ($\alpha \approx 0.2$ in Eq.~\eqref{eq:nuumain}), 4MOST will improve on the sky coverage, which will reach roughly 17000 deg$^2$, and on the number of targets, expected to be of order $\approx 450 \times 10^3$ with redshift depth $z\sim 0.15$. This survey is highly complementary to DESI in that it covers mostly non-overlapping parts of the sky, and so these to datasets could be combined to prove an area of order of 30000 deg$^{2}$. 
\item[LSST]  The Legacy Survey of Space and Time (LSST)\footnote{\href{https://www.lsst.org/about}{https://www.lsst.org/about}} is expected to be operational and running by 2024 at the Vera C. Rubin Observatory~\cite{LSST:2008ijt}, currently under construction in Chile. Over it's 10-year program, it is expected to detect millions of Type Ia supernovae, of which some small fraction will have sufficient light-curve measurements (from LSST or other follow-up campaigns) and host galaxy redshifts/properties to be cosmologically useful. Following ~\cite{Howlett:2017asw}, we consider a hypothetical 10-year survey consisting of $\approx 160 \times 10^3$ galaxies hosting Type Ia Supernovae with threshold J-band magnitude $J < 19$, over a an area of $\sim 18000$ deg$^2$ and a redshift depth of order $z\sim 0.5$. Such a number of objects may be feasible to obtain given already planned or ongoing spectroscopic surveys within the LSST footprint (for instance 4MOST mentioned previously) Although relatively rare, the benefit of using Type Ia supernoave peculiar velocities is that the uncertainty is much better, scaling as 5-10$\%$ of the distance to the target. In this work we use an optimistic (but not overly so) value of $\alpha = 0.05$ in Eq.~\eqref{eq:nuumain}.
\end{enumerate}
\subsection{Reconstructed PV}

As discussed above, ongoing and future peculiar velocity surveys such as DESI, 4HS, and LSST will be an order of magnitude larger than previous surveys. However, with the current available methods for direct PV measurements (e.g. Tully-Fisher, Fundamental Plane, and SN~Ia) the typical uncertainty scales as a considerable fraction of the distance. Thus, beyond  redshift $z\approx0.1$, the uncertainty can become an order of magnitude bigger than the peculiar velocity itself, the latter being typically of order a few hundred km s$^{-1}$. This stands as a stumbling block in extending the PV surveys to higher redshifts, where the overlap with the CMB lensing kernel, and hence the signal in $C^{uk}_{\ell}$ is largest.

Luckily it is possible to to infer indirectly the PV field from the underlying density field. Indeed, in the linear regime the peculiar velocity is attributed to the growing clustering of the total matter (both dark and luminous matter) via Eq.~\eqref{linearvel}, which can be integrated to reconstruct the PV field. This equation however suffers a few limitations: first, we can only measure the galaxy density $\delta_g$ and not the total density $\delta$; second, PV are sensitive to scales beyond the redshift survey limit where we have no direct information. We can overcome these obstacles by assuming that linear biasing holds, $\delta = \delta_g/b$ where $b$ is the linear biasing parameter, and adding an extra free parameter \Vext~to Eq. \ref{linearvel}, accounting for the structures outside the survey limits. We can then readily integrate Eq.\eqref{linearvel} to get the PV field as a function of \Vext~ and $b$. These parameters can be calibrated using a small number of directly measured peculiar velocities, providing a reconstructed PV map. This method has a long history, see for example Refs.\cite{Davis:1996ir,Davis:2010sw,1991ApJ...372..380Y}, and was recently  used by \cite{Carricketal2015} to reconstruct the velocity field in the local universe using positions of $\approx 70000$ galaxies calibrated with $\approx 3000$ directly observed PVs. The final map depends on the smoothing scale chosen for the reconstruction, and Ref.~\cite{Carricketal2015} concludes that a smoothing length of 4 Mpc $h^{-1}$ gives the best peculiar velocity uncertainty of 250~km s$^{-1}$ for the field galaxies and 150~km s$^{-1}$ for groups.
The assumption of a constant full-sky noise for reconstruction is clearly unrealistic, given the local environmental dependence of the noise in the density map from which the reconstruction is derived. On the other hand, the choice of a constant dispersion of 250~km s$^{-1}$ is a conservative one, if compared for example with the average dispersion obtained from the space-dependent noise map of Ref.\cite{Courtois:2022mxo} (which includes the volume covered by \cite{Carricketal2015}).Thus, even if slightly unrealistic, we expect our constant noise map to overestimate the typical errors in the reconstruction, making the conclusions we draw from it robust.   

In this work we assume that their estimation of 250~km s$^{-1}$ with a smoothing scale of 4 Mpc $h^{-1}$ can be extrapolated up to redshift $z<0.15$ covering the full sky. Redshift surveys covering this volume are expected to be available in the near future from surveys such as DESI, 4HS and others not mentioned above. Thus our idealized reconstructed PV catalog contains $\approx 1.6\times10^{7}$ objects (implying $\bar{n} = 1.2\times10^{6}$) with a constant dispersion $\sigma_{rec} = 250$~km s$^{-1}$ replacing the factor $\alpha H_0 \chi \rightarrow \sigma_{rec}$ in Eq.~\eqref{eq:nuumain}.

\subsection{CMB lensing catalogs}
\begin{enumerate}
    \item[PL18] The European Space Agency's \textit{Planck} spacecraft\footnote{\url{https://www.cosmos.esa.int/web/planck}} operated from 2009 to 2013 and performed full-sky measurements of the CMB intensity and polarization anisotropies across nine frequencies between 30 and 857 GHz with the High Frequency Instrument (HFI) and the Low Frequency Instrument (LFI). The final data were released in 2018, and contain the lensing potential map reconstructed from foreground-cleaned CMB temperature and polarization maps~\cite{Planck:2018lbu}. The CMB lensing convergence map covers roughly $70\%$ of the sky (we assume an area of 27500 deg$^{2}$ in this work). The S/N forecasts below are obtained using the true CMB lensing noise curve measured from the data and available on the \textit{Planck} Legacy Archive.\footnote{\url{http://pla.esac.esa.int/pla/\#cosmology}}

    \item[SO] The Simons Observatory is a new-generation ground-based array of millimeter-wave telescopes currently under construction in the high Atacama Desert of Chile \citep{SO} with first light planned for 2023. It will be equipped with a 6-meter Large-Aperture Telescope (LAT) and three smaller telescopes (SATs) and will image the CMB anisotropies over about 40\% of the sky (16500 deg$^{2}$) in six frequencies between 27 and 280 GHz. To predict the feasibility of a possible detection of the cross-correlation signal, we use the official CMB lensing noise curve\footnote{\url{https://github.com/simonsobs/so_noise_models/tree/master/LAT_lensing_noise/lensing_v3_1_1}} released by the SO collaboration calculated assuming foreground cleaned CMB maps and using CMB multipoles between $30 \le \ell \le 3000$ in temperature and  $30 \le \ell \le 5000$ in polarization.
    
    \item[CMB-S4] The fourth-generation ground-based CMB experiment, CMB-S4, is currently in its design stages and expected to start operations later this decade \citep{CMB-S4}. In its current configuration, 21 telescopes between the South Pole and the Atacama Desert in Chile will survey roughly 70\% of the sky (again, in this work we assume 27500 deg$^{2}$). Relevant to this work, the LAT survey will use 6-metre class telescopes in six frequency bands from 30 to 270 GHz. Adopting the individual frequency noise levels from \cite{Bianchini:2022wte}, we calculate the CMB lensing reconstruction noise curve from component-separated CMB using temperature and polarization information (in the $30 \le \ell \le 3000$ and  $30 \le \ell \le 5000$ ranges respectively).
\end{enumerate}


\section{Results}
\label{Results}
\begin{figure}[h]
\includegraphics[scale=0.8]{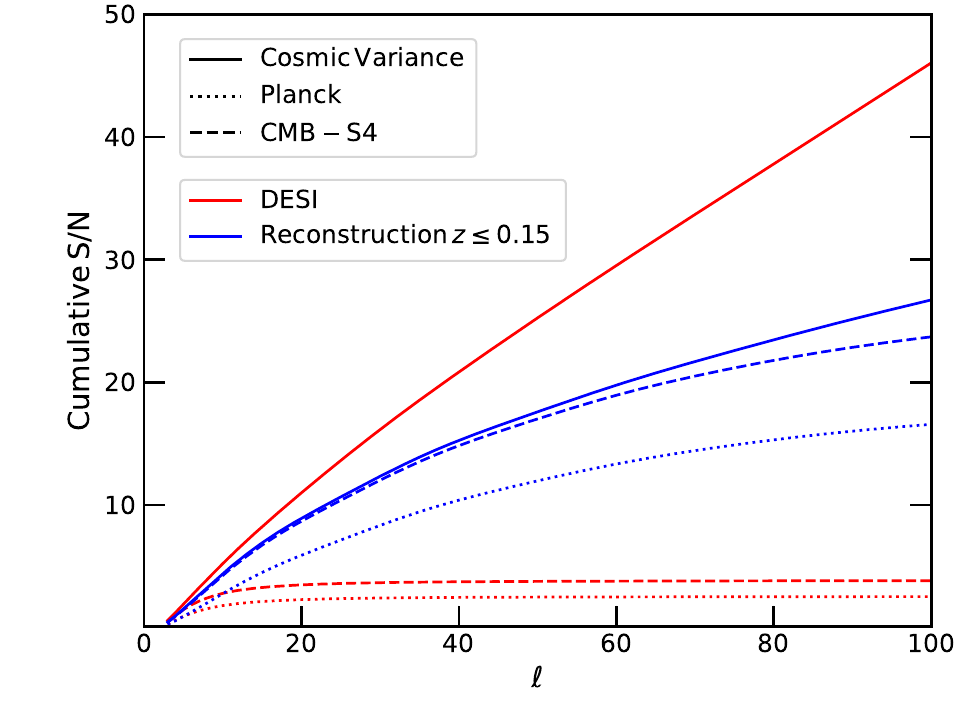}
	\caption{The cumulative signal-to-noise ratio, with $\ell \geq 3$, as a function of maximum $\ell$-mode for different representative combinations of PV and CMB dataset. The solid lines show the sample variance limit --- the case where the shot/instrumental noise in both the CMB lensing convergence and peculiar maps is assumed to be zero.}
	\label{SN}
\end{figure}

The main results of this work, i.e. the cumulative Signal-to-Noise ratio (S/N) at angular scales $ 3\leq \ell \leq 200$ for different PV and CMB surveys are reported in Fig.~\ref{SN} and in Table~\ref{TableSN}. In all cases, we assume complete overlap between the PV and CMB lensing surveys, and take the smallest area of the two when computing the sky fraction $f_{sky}$. The biggest angular scale that can be be probed by all the surveys considered in this work corresponds to a lower bound on the the angular mode $\ell = 3$, which allows for a fairer comparison of the cumulative S/N from different surveys.
It is clear from Fig.~\ref{SN} that the cumulative S/N starts to plateau at small scales, i.e. large multipole $\ell$. One cause for this is due to the lack of intrinsic signal at these scales, which can be seen by looking at Fig.~\ref{Cells}, and is as expected by virtue of Eq.~\eqref{CCPVCMBLimber} which explicitly shows that $C^{u\kappa}_\ell$ scales with $\ell^{-1}$. This in turns implies that a more conservative choice of angular scales without the first low-$\ell$ modes results in a significant degradation of the S/N, as one can appreciate comparing the upper and lower panels of table~\ref{TableSN}. We dedicate the rest of this section to a more thorough discussion of our results, particularly focused on the impact of the PV errors.

\begin{table}
\centering
\begin{tabular}{|c|ccc|}
\hhline{|====|}
\multirow{2}{*}{\diagbox{PV Survey}{CMB Survey}} & Planck & SO & CMB-S4 \\
& ($27500\,\mathrm{deg}^{2}$) & ($16500\,\mathrm{deg}^{2}$) & ($27500\,\mathrm{deg}^{2}$) \\ \hline
DESI ($14000\,\mathrm{deg}^{2}$) & 2.5 & 3.6 & 3.8 \rule{0pt}{2.6ex} \\
4HS ($17000\,\mathrm{deg}^{2}$) & 2.5 & 3.5 & 3.8 \\
LSST ($18000\,\mathrm{deg}^{2}$) & 2.4 & 3.2 & 3.5 \\
Reconstruction $z\leq0.15$ (Full Sky) & 18.7 & 20.1 & 27.4 \rule[-1.2ex]{0pt}{0pt} \\
\hhline{|====|}
\end{tabular}

\bigskip

\centering
\begin{tabular}{|c|ccc|}
\hhline{|====|}
\multirow{2}{*}{\diagbox{PV Survey}{CMB Survey}} & Planck & SO & CMB-S4 \\
& ($27500\,\mathrm{deg}^{2}$) & ($16500\,\mathrm{deg}^{2}$) & ($27500\,\mathrm{deg}^{2}$) \\ \hline 
DESI ($14000\,\mathrm{deg}^{2}$) & 1.2 & 1.6 & 1.7 \rule{0pt}{2.6ex} \\
4HS ($17000\,\mathrm{deg}^{2}$) & 1.5 & 2.0 & 2.1 \\
LSST ($18000\,\mathrm{deg}^{2}$) & 1.9 & 2.5 & 2.7 \\
Reconstruction $z\leq0.15$ (Full Sky) & 17.0 & 18.2 & 24.8 \rule[-1.2ex]{0pt}{0pt} \\
\hhline{|====|}
\end{tabular}

\caption{Predicted cumulative signal-to-noise ratios in $C_{\ell}^{u \kappa} $ for $3 \leq \ell \leq 200$ (upper table) and $8 \leq \ell \leq 200$ (lower table) for different combinations of peculiar velocity (PV) and CMB-lensing experiments. The distribution and the number of PV sources in each survey are given in Fig.~\ref{windows}. For each combination of experiments, we treat the overlap area as the minimum of the two PV and CMB surveys.
}
\label{TableSN} 
\end{table}

\subsection{Cross Correlation using direct PV measurements}

\begin{figure}[h]
\centering

\includegraphics[scale=0.7]{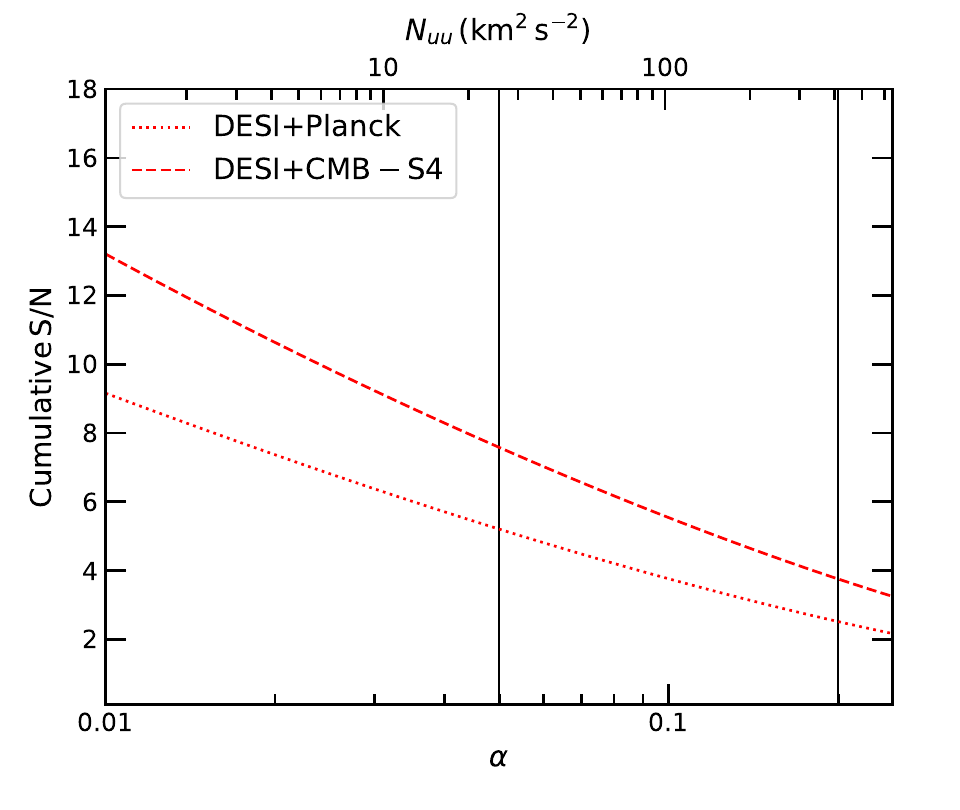}
	\caption{The cumulative S/N for $C^{u\kappa}_\ell$ for a DESI-like survey as a function of $\alpha$, the typical uncertainty of PV measurenents as a fraction of $H_0\chi$.}
	\label{SNalpha}
\end{figure}

We see from Table~\ref{TableSN} that in the best case scenario, obtained by combining the direct PV measurements with the convergence map forecast for CMB S4, the cumulative S/N is above the 3$\sigma$ threshold and therefore we could potentially marginally detect the cross correlation signal  (see the dashed red line in Fig.~\ref{SN}). 
On the other hand, if we consider idealized PV surveys with no error (so that the only source of uncertainty is the sample variance), the cumulative S/N becomes greater than $\geq 10$ already at relatively small $\ell\approx 20$ (see the solid red line in Fig.~\ref{SN}). 
One can identify the causes of such a degradation of the S/N ratio by looking at the terms entering the covariance matrix of Eq.\eqref{covar} in  Fig.~\ref{Cells}. We see, comparing the solid and dotted lines in the upper right panel, that the Planck noise is always comparable or bigger than the intrinsic signal, whereas the expected noise from SO and CMB S4 is always smaller. This explains the difference between the first and the other columns in Table~\ref{TableSN}. On the other hand, by looking at the upper left panel of Fig.~\ref{Cells}, we easily realize that the largest source of uncertainty comes from the intrinsic dispersion of PV measurements, for which the noise is bigger than the signal's theoretical prediction already for $\ell \geq 5$.
This noise is largely dominated by the uncertainty on the position of the source, usually derived from scaling relations such as the Fundamental Plane (FP) or the Tully-Fisher (TF) ones. As a result, the dispersion on direct PV measurements usually scales with the distance to the source, leading to a quick degradation of the S/N. This explains the earlier flattening of the dashed and dotted red curves in Fig.~\ref{SN} compared to the blue ones.
In Fig.\ref{SNalpha} one can appreciate how the S/N changes as a function of $\alpha$ for a DESI-like survey combined with Planck or CMB-S4.

\subsection{Cross correlation using reconstructed PV}
\begin{figure}[h]
\centering

\includegraphics[scale=0.7]{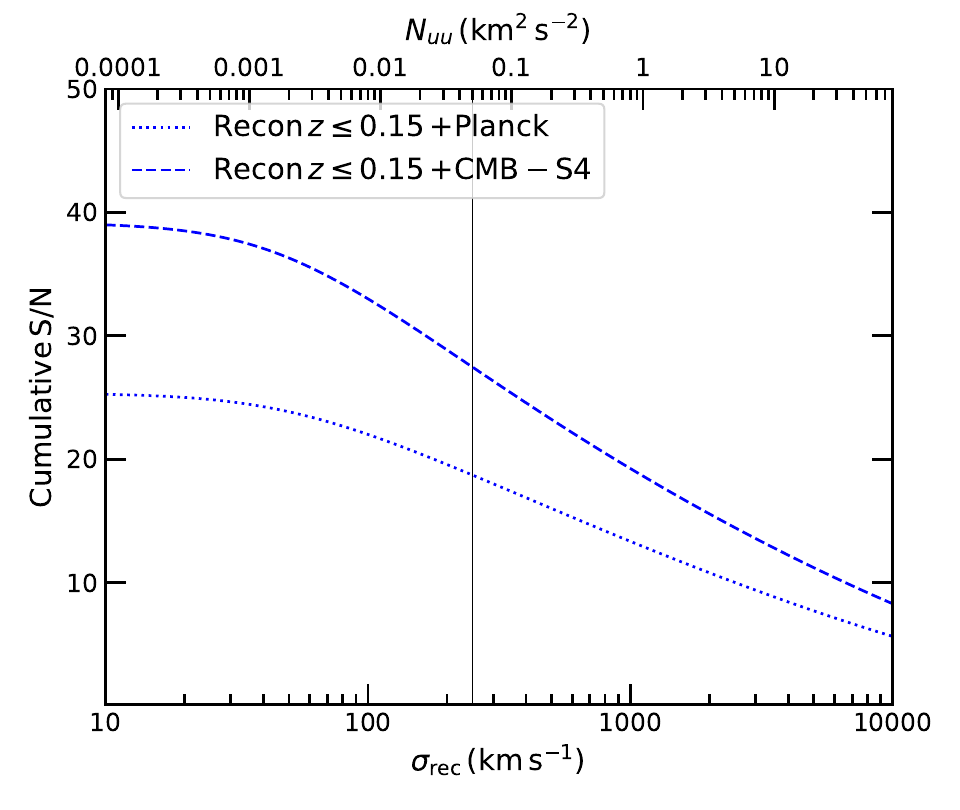}
	\caption{The cumulative S/N for $C^{u\kappa}_\ell$ derived from a reconstructed PV map extending up to redshift $z\leq 0.15$ as a function of the value of the assumed constant dispersion $\sigma_{rec}$.}
	\label{SNsigmarec}
\end{figure}

We know that at redshift $0\leq z\leq 1$, the lensing kernel grows monotonically (see for example the dashed line of Fig.~\ref{windows}), hence the cross correlation signal should increase with PV measurements at higher redshifts.  On the other hand, as previously discussed, the noise associated to direct PV measurements scales with the distance and causes a worsening of the overall S/N. To understand how this picture would change with PV dispersions without such scaling, we studied the cross correlation with a PV map obtained through a reconstruction technique, for which the dispersion is a constant determined by the choice of the reconstruction smoothing scale. For an idealized reconstructed PV field which extends up to redshift $z= 0.15$, with a smoothing scale of $4$ Mpc $h^{-1}$ and a constant dispersion $\sigma_{rec}= 250$~km s$^{-1}$ we see from Fig.~\ref{SN} that the cumulative S/N improves greatly and approaches the sample variance limit. 
To estimate the dependence of the signal from the constant dispersion assumed for the reconstructed map we plot in Fig.~\ref{SNsigmarec} the cumulative S/N as a function of the value of the constant dispersion $\sigma_{rec}$, from which we see that the cumulative S/N is still $\geq 10$ even if the constant noise is of order $\sigma_{rec} \sim 10^3$. 
It is important to stress that the reconstructed PV map depends somehow on the galaxy density field and the galaxy bias $b(z)$, and therefore cannot be considered as a completely independent tracer of the underlying DM distribution. However, this exercise also shows the potential of $C^{u\kappa}_\ell$ as an unbiased cosmological probe if direct PV measurements with the precision of the reconstructed ones will be available in the future. Some examples could be through the construction of direct PV surveys with higher number density than we have considered here, the use of ancillary properties such as metallicity or age in the standard Tully-Fisher or Fundamental Plane relations to reduce their intrinsic scatter \cite{Magoulas,Ouelette,deugenio}, the use of ‘twinned’ supernovae to obtain more accurate distance moduli \cite{Fakhouri}, or alternate distance indicators such as Brightest Cluster Galaxies \cite{Lauer}, Type-II supernovae \cite{Stahl:2021mat}, or even gravitational waves\cite{Palmese:2020kxn}. However, while promising, more work is needed in this area to realise the potential of, and implement, these improvements in future surveys.

\section{Conclusions and outlook}
\label{Discussion}
Both the peculiar motion of individual galaxies and the optical path of CMB photons are influenced by the gravitational field of the cosmic web, and thus can be used as DM tracers. Their statistical properties can be studied together with the distribution of galaxies in the sky. On the other hand, contrary to the galaxy distribution, they have no dependence from the galaxy bias $b(z)$, and thus can be used to break degeneracies in our cosmological inference.\footnote{It is important to stress that this is true for directly measured peculiar velocities, but not for the ones obtained through reconstruction as they require an underlying galaxy density field to be defined.} Furthermore, PV's are explicitly sensitive to the growth rate $f$, whereas the lensing convergence to the matter distribution $\Omega_m$, making them a promising tool to explore different gravitational models and current cosmological parameters tensions. In this work, for the first time, we assess the significance of their statistical cross correlation from current and forthcoming experiments. Because of the geometry of the problem, one would expect the signal to be intrinsically small.
Indeed, the CMB lensing peaks at relatively high redshift $z \sim \mathcal{O}(1)$, whereas peculiar motions observations are available only at closer redshifts $z \sim \mathcal{O}(0.1)$. However, contrary to these expectations, our theoretical prediction shows that with perfect experiments, i.e. at the sample variance limit, the signal would be significantly detectable with a cumulative S/N $\geq 30 \sigma$ (standard deviations) at angular scales $3\leq\ell\leq100$. If realistic noises are taken into account, however, the S/N drops for direct PV measurements to a few $\sigma$'s, making their detection marginal.

Looking at Eq.~\eqref{covar} and Fig.~\ref{Cells} it is straightforward to realize that the observed PV uncertainties are responsible for such a degradation of the cumulative S/N, since the noise starts to dominate over the signal already at relatively large scales (small $\ell$).
This is due to the fact that the error on the source's PV is usually proportional to its comoving distance to the observer, and thus increases with the redshift.
Another option is to use a reconstructed peculiar velocity map, for which the uncertainty is approximately constant with the redshift, depending on the choice of the reconstruction smoothing scale. For a reconstruction that extends up to redshift $z\leq 0.15$, a smoothing scale of $4h^{-1}$ Mpc (i.e. one likely to be feasible with DESI and 4HS data) and CMB S4 observations, the dashed blue line in Fig.~\ref{SN} shows that the cumulative S/N approaches the sample variance limit. Whilst a reconstructed PV map would not be completely independent of the galaxy bias $b(z)$, the above exercise shows the expected significance of such a cross-correlation if the errors on PV observations do not grow with redshift.

In summary, this paper shows that the radial peculiar velocities of galaxies at low redshift and the convergence map of CMB photons on angular scales $\mathcal{O}(1) \leq \ell\leq \mathcal{O}(100)$ are correlated, and such correlation might be detectable with forthcoming surveys in the near future. It would be interesting to assess the constraining power of such a probe for cosmological inference of $\Lambda$CDM derived parameters like the growth rate $f$ or the  Hubble factor today $H_0$, as well as for testing different theories of gravity and cosmology beyond the standard model. We will address these compelling questions in future works.

\section*{Acknowledgement}
We are grateful to Hayley Valiantis for their support during the preparation of this paper. Part of the work was  financially supported by the Australian Government through the Australian Research Council Laureate Fellowship grant FL180100168. 

\appendix
\section{Limber approximation for $C^{u\kappa}_\ell$}\label{appendixA}
The spherical Bessel functions of order $n$ may be written using
\begin{equation}\label{Besseldef}
    j_n(x) = \left(-x\right)^n \left(\frac{1}{x}\frac{d}{dx}\right)^n\frac{\sin{x}}{x}\;.
\end{equation}
By taking derivatives of both sides of this equation one can explicitly check that the following differential relation holds:
\begin{equation}\label{diffBess}
    \left(\frac{1}{x}\frac{d}{dx}\right)^m\left(x^{n+1} j_n(x)\right)=x^{n-m+1}j_{n-m}(x)\;.
\end{equation}
Using the above we can rewrite the the first derivative of the spherical Bessel function as
\begin{equation}
    j'_n(x)= j_{n-1}(x) -\frac{n+1}{x}j_n(x)\;,
\end{equation}
which allows us to rewrite Eq.~\eqref{CCPVCMB} as the sum of two terms:
\begin{equation}\label{almostLimber}
    C_\ell^{u\kappa} \equiv \frac{2}{\pi}\int dk \;d\chi_1\;d\chi_2\;W^{u}(\chi_1)W^{\kappa}(\chi_2)\left[j_{\ell -1}\left(k\chi_1\right) -\frac{\ell+1}{k\chi_1}j_{\ell}\left(k\chi_1\right)\right] j_\ell\left(k\chi_2\right)P_m\left(k,\chi_1,\chi_2\right)\;.
\end{equation}
We can further simplify the above expression by making use of the Limber approximation, which  evaluates integrals containing products with the spherical Bessel functions $j_\ell(x)$ according to
\begin{equation}\label{Limberform}
\lim_{\epsilon \rightarrow 0} \int dx\;\sqrt{\frac{2x}{\pi}} e^{\epsilon\left(x - \ell\right)} j_{\ell-\frac{1}{2}}(x)f(x)  \approx f(\ell) +\mathcal{O}\left(\ell^{-2}\right)\;,    
\end{equation}
where the magnitude of the $\ell^{-2}$ correction depends explicitly on the position and the width of the peaks of the window functions,
see for example Ref.\cite{LoVerde:2008re} for a formal derivation. Eq.~\eqref{CCPVCMBLimber} is then easily obtained using Eq.\eqref{Limberform} in Eq.~\eqref{almostLimber} to approximate the integrals over $d\chi_1$ and $d\chi_2$. 

\section{Shot noise for the projected radial peculiar velocity field}\label{appendixb}

The effect of noise in the radial peculiar velocity field can be computed following the same derivation as for any observable in Section~\ref{angcorr}. If we define the observed peculiar velocity field as $\tilde{u}(\boldsymbol{\hat{n}}, \chi) = u(\boldsymbol{\hat{n}}, \chi) + \xi_{u}(\boldsymbol{\hat{n}}, \chi)$ then the observed angular velocity-velocity power spectrum can be written $\tilde{C}_\ell^{uu} = C_\ell^{uu} + N_\ell^{uu}$ where $N_\ell^{uu}$ denotes the contribution from the noise in the velocity field
\begin{equation}
N_\ell^{uu} = \int d\Omega_{\hat{\textbf{n}}} d\Omega_{\hat{\textbf{n}}'}\; Y^{*}_{\ell m}(\hat{\textbf{n}})Y_{\ell ' m'}(\hat{\textbf{n}}') \langle \xi_{u}(\boldsymbol{\hat{n}}, \chi_{1}) \xi_{u}(\boldsymbol{\hat{n}}', \chi_{2})\rangle.
\end{equation}
In practice, the angular power spectrum is computed by projecting the observable over the redshift bin. For the radial peculiar velocity this projection kernel is $dn/d\chi$. This act of projection also applies to the noise spectrum, such that
\begin{equation}
N_\ell^{uu} \Rightarrow N_\ell^{uu} = \int \frac{dn}{d\chi_{1}} d\chi_{1} \int \frac{dn}{d\chi_{2}} d\chi_{2} \int d\Omega_{\hat{\textbf{n}}} d\Omega_{\hat{\textbf{n}}'}\; Y^{*}_{\ell m}(\hat{\textbf{n}})Y_{\ell ' m'}(\hat{\textbf{n}}')\langle \xi(\boldsymbol{\hat{n}}, \chi_{1}) \xi(\boldsymbol{\hat{n}}', \chi_{2}) \rangle
\label{eq:nuu}
\end{equation}

As discussed in \cite{Koda:2013eya, Howlett:2017asw, Howlett:2017asq, Howlett:2019bky} the error in the radial peculiar velocity field generally scales as $\xi_{u}(\boldsymbol{\hat{n}}, \chi) = \alpha H_{0} \chi / \sqrt{N}$ where the $1/\sqrt{N}$ term arises from the averaging of $N$ independent radial peculiar velocity measurements and typically $\alpha=0.2$ for Tully-Fisher or Fundamental Plane distance measurements and $\alpha=0.05$ for Type Ia Supernovae. Crucially, the errors typically depend on the distance to the object, but not on the direction on the sky. Hence, if we substitute this expression into Eq.~\ref{eq:nuu} we can integrate out the spherical harmonics
\begin{align}
N_\ell^{uu} &= \frac{1}{N} \int d\Omega_{\hat{\textbf{n}}} d\Omega_{\hat{\textbf{n}}'}\; Y^{*}_{\ell m}(\hat{\textbf{n}})Y_{\ell ' m'}(\hat{\textbf{n}}')\int d\chi_{1} \frac{dn}{d\chi_{1}} \alpha H_{0} \chi_{1} d\chi_{1} \int \frac{dn}{d\chi_{2}} \alpha H_{0} \chi_{2} d\chi_{2} \notag \\
&= \frac{1}{\bar{n}}\left(\int \frac{dn}{d\chi} \alpha H_{0} \chi d\chi\right)^{2},
\end{align}
where the resulting $4\pi$ has been used to convert the number of galaxies to an angular number density $\bar{n}$. This expression matches that in Eq.~\ref{eq:nuumain} from the main text.

\bibliographystyle{unsrturl}
\bibliography{References.bib}
\end{document}